# Phase-field modeling of two-dimensional solute precipitation/dissolution: Solid fingers and diffusion-limited precipitation

Zhijie Xu[1,a)] and Paul Meakin[2,3,4]

1. Computational Mathematics Group, Fundamental and Computational Sciences Directorate, Pacific Northwest National Laboratory, Richland, WA 99352, USA

2. Center for Advanced Modeling and Simulation, Idaho National Laboratory, Idaho Falls, Idaho 83415, USA.

3. Physics of Geological Processes, University of Oslo, Oslo 0316, Norway

4. Multiphase Flow Assurance Innovation Center, Institute for Energy Technology, Kjeller 2027, Norway

Two-dimensional dendritic growth due to solute precipitation was simulated using a phase-field model reported earlier [Z. Xu and P. Meakin, J. Chem. Phys. 129, 014705 (2008)]. It was shown that diffusion-limited precipitation due to the chemical reaction at the solid-liquid interface posses similarities with diffusion-limited aggregation (DLA). The diffusion-limited precipitation is attained by setting the chemical reaction rate much larger compared to the solute diffusion to eliminate the effect of the interface growth kinetics. The phase-field simulation results were in reasonable agreement with the analytical solutions. The fractal solid fingers can be formed in the diffusion-limited precipitation and have a fractal dimension measured $d_f = 1.68$, close to 1.64, the fractal dimensionality of large square lattice diffusion-limited aggregation (DLA) clusters.

a) Electronic mail: zhijie.xu@pnl.gov

## I. INTRODUCTION

Patterns and pattern formation have fascinated humanity for centuries, and they provide important clues about the materials and processes that form them. Many systems that are important for scientific and practical reasons form dendritic patterns (rivers, trees, blood vessels, microorganism colonies and crystalline mineral and metallic dendrites are conspicuous examples). The growth of dendritic structures has been investigated by physicists, chemists and biologists for centuries. In 1611, Kepler wrote a manuscript[1] that related the six-fold symmetry of snowflakes to the packing of spheres. The complex forms and varied shapes of snowflakes can best be explained by the idea that the growth process is sensitive to the changing environment (temperature, water supersaturation, impurities and etc.) in which the snowflake is growing. The distance between the six arms of the snowflake is small enough to ensure that all six arms experience essentially the same changing growth conditions, but too large for the growth of one arm to be directly controlled by the growth of its neighbors. The six-fold symmetry is clearly related to the internal crystal structure, as Kepler suspected but could not prove. In practice, the sensitivity to growth conditions and random atomic scale events is so great that the six-fold symmetry that makes snowflakes remarkable is far from perfect. This can be attributed to the Mullins-Sekerka instability,[2] which amplifies shape perturbations that occur during the growth process. The Mullins-Sekerka instability is also responsible for the dendritic form of the snowflake. Without this instability, the growth of ice in cold moist air would lead to the formation of compact crystals with faceted (euhedral) shapes. The beautifully written book titled "*On growth and form*" by D'Arcy Wentworth Thompson,[3] marked another milestone in the scientific investigation of pattern formation, and during recent decades simple computer models have played an important role in the development of a better understanding of pattern formation processes.

The DLA model,[1-4] originally introduced by Witten and Sander[4] to explain irreversible particle aggregation, is relevant to a wide range of growth processes that result in the formation of dendritic patterns. Important examples include viscous fingers,[5] dielectric breakdown,[6] the formation of a variety of branched biological structures and the chemical dissolution of a porous medium.[7]

Precipitation processes play an important role in the formation of a wide range of geological patterns such as stalactites stalagmites, travertine terraces, faceted crystals and mineral dendrites. Dissolution and precipitation change pore space and fracture aperture geometries, resulting in changes in porosity and permeability, and this may have important consequences in geotechnological applications such as oil recovery, carbon dioxide sequestration and the management of chemical and radioactive waste. In general, dissolution and precipitation are controlled by the transport of dissolved substances due to a combination of advection and diffusion and chemical processes at the moving solid-liquid interface. The kinetics of precipitation and/or dissolution depends on the orientation of the surface with respect to the crystallographic axes (the Miller indices), microscopic details such as the populations of various defects, including both structural defects and impurities, and in some cases, diffusion along the solid/liquid interface.

Pattern formation resulting from dissolution and precipitation can be understood in terms of the dynamics of the solid-liquid boundary, and the problem of simulating this class of pattern formation processes is equivalent to solving the corresponding moving boundary problem or Stefan problem. Moving boundary problems have a reputation for being difficult to solve numerically, and substantial effort has been made to develop fast and reliable methods. This has resulted in the introduction of front-tracking,[8] volume-of-fluid,[9] level set[10] and other methods during the past two decades. These methods have relative strengths and weaknesses, and the selection of the interface tracking/capturing method depends on the needs of the particular application. For example, the volume of fluid method is based on a relatively crude representation of the moving boundary (line segments in each grid cell, which do not join those in adjacent cells), but mass conservation is good. On the other hand, the level set method is based on a continuous boundary representation that allows the boundary to be located with subgrid scale resolution, but mass conservation is poor in simple level set method. The phase-field method, which we used in the work reported here, is based on the diffuse-interface theory.[11, 12] The advantage of this approach is that there is no need to explicitly track the interface, and toplogical changes such as coalescence and fragmentation can be simulated in a robust and reliable manner. This makes the phase-field approach attractive for simulating precipitation/dissolution processes in systems with complicated geometries where topological changes may occur. In diffuse-interface models (grid-based phase-field method or the hybrid diffuse-interface/particle method we recently developed[13]), physical properties are always assumed to vary smoothly with position from one phase to the other across a thin interface zone with a non-zero width, $w$. Beginning with applications to the solidifications of pure melts,[14] the phase-field approach has been used in a wide range of applications including solidification coupled with melt convection,[15] two-phase Navier-Stokes flow,[16] and solid precipitation and dissolution.[17] Due to its subgrid scale accuracy, the phase-field method was also recently employed for an efficient implementation of no-slip boundary conditions in particle multiphase fluid flow models and applications (for example, dissipative particle dynamics and smoothed particle hydrodynamics).[18, 19]

In a variety of dendritic growth processes diffusion is the dominant solute transport process, and the diffusion-limited aggregation (DLA) provides important insight. In the original DLA model, a lattice site is filled to represent a seed or nucleus for the growth process. A random walk is then started far from the seed, and when the random walker reaches a site on the perimeter of the seed that site is filled. The process of starting a random walk far from the cluster of filled sites, filling the first site on the perimeter of the cluster reached by the random walk and starting a new random walk far from the growing cluster is then repeated many times until a large cluster of sites is formed. A variety of modifications are used to make this algorithm practical. For example, the random walk is started at a random point on a circle centered on the seed that just encloses the cluster and moved to the nearest lattice site, and if the random walk moves too far from the growing cluster, a new random walk is started. The clusters grows irreversibly to form a random dendritic strutures.[20] Precipitation in the absence of fluid flow is similar to diffusion-limited aggregation because solute is transported to the solid-liquid interface and diffusion is the only mechanism responsible for solute transport. However, the kinetics of the incorporation of solute into the solid phase at the interface is also important, and if this kinetics is slow enough, the process will become reaction-limited, and the precipitate will be compact instead of dendritic. In some applications, such as the formation of single crystal dendrites by growth from a melt or precipitation from solution, the DLA model generates branched structures that are too disorder. The disorder can be reduced by a noise reduction algorithms,[21] and a variety of other modifications can be made to the DLA model for specific applications.[22]

Here we use a phase-field model developed recently[17] for solid precipitation processes to investigate the continuum counterpart of the DLA model. In the continuum model the interface must be assumed to be a perfect absorber for the solute (the limit $k \to \infty$ is assumed where $k$ is the rate constant for the precipitation reaction). In general, the growth of a solid body by precipitation from a dilute supersaturated solution is an unstable process due to the Mullins Sekerka instability. Using a linear stability analysis, Mullins and Sekerka[2, 23] showed that for a spherical particle in a diffusion-cotrolled growth into supersaturated matrix where a sinusoidal perturbation is introduced with a wavelength of $\lambda$ or a wavenumber of $k_\lambda = 2\pi/\lambda$, the growth velocity is given by

$$V(k_\lambda) = v_0 k_\lambda (1 - B\xi_C \xi_D k_\lambda^2), \tag{1}$$

where $\xi_D$ is the diffusion length, $\xi_C$ is the capillary length, $v_0$ is the average growth velocity and $B$ is a constant of order unity. The diffusion length is given by $\xi_D = D/v_0$, where $D$ is the diffusion coefficient, and the capillary length is given by

$$\xi_C = C_e^\infty \gamma V_m / \Delta C k_B T \tag{2}$$

where $\gamma$ is the interfacial energy, $k_B$ is the Boltzmann constant, $V_m$ is the molecular volume in the solid phase and $\Delta C = C_\infty - C_e^\infty$, where $C_\infty$ is the far-field solute concentration and $C_e^\infty$ is the solute concentration in equilibrium with a flat solid surface. In the model that we investigate in this work, the capillary length given by Eq. (2) is zero, because there is no surface tension. Under these conditions, Eq.(1) indicates that the growth rate of the perturbation increases linearly with increasing wavenumber, $k_\lambda$. However, because of the limited resolution of the lattice based model used in this work, represented by the size, $a$, of the lattice grid elements, the dispersion function, $V(k_\lambda)$ is cut off at a scale of the order of $a^{-1}$, as Fig. 1 shows. Consequently, the wavelength of maximum growth (the characteristic morphology length) is of the order $a$.

## II. SHARP-INTERFACE AND PHASE-FIELD EQUATIONS

The phase-field model used in this work is based on the sharp-interface model of solute precipitation and/or dissolution. The simplest model for solute precipitation/dissolution includes diffusion in the liquid and a reaction with first order kinetics at the liquid-solid interface in the absence of fluid flow. The dynamics of the solid-liquid interface during dissolution or precipitation is a result of the transport of dissolved solid from or to the interface (diffusion in the solid phase is usually small enough to neglect). The system of sharp-interface equations for precipitation and/or dissolution is:

$$\partial c / \partial t = D \nabla^2 c, \qquad (3)$$

$$c\big|^+ = v_s / bkk_c \text{ on } \Gamma, \qquad (4)$$

$$v_s = bk_c k \, c\big|^+ = bk_c D \nabla c\big|^+ \cdot \mathbf{n} \text{ on } \Gamma, \qquad (5)$$

where $D$ is the solute diffusion coefficient. $c = (C - C_e^\infty)/C_e^\infty$ is the normalized solute concentration ($C$ is the solute concentration, and $C_e^\infty$ is the solute concentration at equilibrium with a flat solid surface). In Eqs. (4) and (5), $\Gamma$ stands for the interface between solid and liquid phases. $c\big|^+$ and $\nabla c\big|^+$ are the dimensionless normalized solute concentrations and concentration gradients at the interface $\Gamma$ with $\big|^+$ indicating the magnitude of a variable at the liquid side of the interface. The dimensionless variable $b$ is defined as $b = C_e^\infty / \rho_s$ ($\rho_s$ is the density of the solid). Equation (5) relates the interface velocity to the solute concentration at the interface (the Gibbs-Thomson effect, which depends on the interfacial free energy density or surface tension, is assumed to be small, and is neglected in the current study), where $\mathbf{v}_s$ is the velocity of the interface in the direction normal to the interface, $k_c$ is a stoichiometric coefficient of order unity,[24] $\mathbf{n}$ is the unit vector perpendicular to the interface pointing into the liquid, and $k$ is the reaction rate coefficient. The system of dimensionless sharp-interface governing equations

$$\partial c/\partial t = \nabla^2 c/P_e,  \quad (6)$$

$$c|^+ = (2\alpha P_e/D_a)v_s \text{ on } \Gamma, \quad (7)$$

$$v_s = (1/2\alpha P_e)\nabla c|^+ \cdot \mathbf{n} = (D_a/2\alpha P_e)c|^+ \text{ on } \Gamma. \quad (8)$$

can be obtained by introducing the units of length $L$, velocity $U$, time $L/U$, and three dimensionless constants, $\alpha = 1/2bk_c$, the Péclet number $P_e = UL/D$ and the Damköhler number $D_a = kL/D$.

In general, it is very difficult to directly solve the sharp-interface equation systems (3)-(5) or (6)-(8). Phase-field method provides a robust approach to solve the sharp-interface equations without explicitly dealing with the quantities at the moving boundary. The original moving boundary problem can be reduced to a set of coupled partial differential equations that is much easier to solve. The equivalent set of phase-field equations to the sharpe-interface Eqs. (3)-(5) have been formulated in detail in our previous work,[17] where an additional variable $\phi(\mathbf{x},t)$ was introduced to indicate which phase is at position $x$ and time $t$ ($\phi$ varies from -1 to 1 with $\phi < -(1-\delta)$ indicating the solid phase and $\phi > 1-\delta$ indicating the liquid phase where $\delta \ll 1$). The phase-field equations are

$$\tau \frac{\partial \phi}{\partial t} = \varepsilon^2 \nabla^2 \phi + \left(1 - \phi^2\right)\left(\phi - \lambda c\right) - \varepsilon^2 \kappa \left|\nabla \phi\right|, \tag{9}$$

$$\frac{\partial c}{\partial t} = D\nabla^2 c + \alpha \frac{\partial \phi}{\partial t}\left(1 + \frac{D\nabla^2 \phi - \partial \phi/\partial t}{k\left|\nabla \phi\right|}\right), \tag{10}$$

$$\tau = \alpha\lambda \frac{\varepsilon^2}{D}\left(\frac{5}{3} + \frac{\sqrt{2}D}{k\varepsilon}\right), \tag{11}$$

where $\tau$, $\varepsilon$, and $\lambda$ are all phase-field microscopic parameters. $\tau$ is a characteristic time parameter. $\varepsilon$ is closely related to the interface thickness and $\lambda$ is a dimensionless parameter that controls strength of the coupling between the phase-field variable, $\phi$, and the concentration field, $c$. The relationship between $\tau$, $\varepsilon$, and $\lambda$ (Eq. (11)) is obtained from a formal asymptotic analysis in reference 17, which ensures that the phase-field solutions to Eqs. (9)-(10) converge to the sharp-interface solutions to Eqs. (3)-(5) in $\varepsilon \to 0$ limit. A comparison between shape-interface and phase-field equations indicates that original interfacial boundary conditions (Eqs. (4) and (5)) do not explicitly appear in the phase-field equations. Instead, Eq. (10) includes two additional source/sink terms on the right hand side that implicitly represent the boundary conditions that must be satisfied at the interface. The first term acts as a net source or sink of solute corresponding to the discontinuity in the solute concentration gradient across the interface, while the second term acts as a net source or sink of solute coming from the discontinuity in the solute concentration across the interface.

For simplicity in the numerical calculations, the phase-field equations (9)-(11) can be further rewritten in a dimensionless form by introducing the units of length $\varepsilon$, velocity $D/\varepsilon$, and time $\varepsilon^2/D$. With these definitions, the new phase-field equations are rewritten as

$$\frac{\partial \phi}{\partial t} = \frac{1}{P_e^{'}}\left\{\nabla^2\phi + \left(1-\phi^2\right)\left(\phi - \lambda c\right) - \kappa^{'}\left|\nabla\phi\right|\right\}, \qquad (12)$$

$$\frac{\partial c}{\partial t} = \nabla^2 c + \alpha\frac{\partial \phi}{\partial t} + \left(\nabla^2\phi - \partial\phi/\partial t\right)\frac{\alpha \cdot \partial\phi/\partial t}{D_a \varepsilon^{'} \left|\nabla\phi\right|}, \qquad (13)$$

$$\lambda = P_e^{'} \Big/ \left[\alpha\left(5/3 + \sqrt{2}/D_a\varepsilon^{'}\right)\right], \qquad (14)$$

$$P_e^{'} = \tau L U \Big/ \left(\varepsilon^2 P_e\right) = \tau D/\varepsilon^2, \; \kappa^{'} = \kappa\varepsilon, \text{ and } \varepsilon^{'} = \varepsilon/L, \qquad (15)$$

where the parameter $P_e^{'}$ is the Péclet number that controls the diffusion of the phase-field variable $\phi$ (Eq. (12)).

Phase-field simulations can be implemented by discretizing Eqs. (12) and (13) on a regular finite difference grid with constant grid spacing $\Delta x$ in both directions. For the two-dimensional square lattice simulations presented here, a grid spacing, $\Delta x$, of $0.25\varepsilon$ to $0.5\varepsilon$ was found to achieve sufficiently accuracy with reasonable computing time,[25] and here $\Delta x = 0.5\varepsilon$ was used for all simulations. Because of the symmetry of the square lattice, simulations were performed on a single quadrant of the lattice, and the 4-fold rotational symmetry was used to obtain calculate the phase field in the other three quadrants and update the phase field, $\phi$, in the grid elements along the boundaries between quadrants. Phase-field simulations were started with a small quarter disk to represent solid with a initial radii $R_0$ ( $R_0 = 5.0\varepsilon$ ) at the lower-left corner of the computational domain (the computational domain size is fixed to be $0.5 \times 0.5$). The far-field concentration boundary condition was approximated by applying a fixed concentration value, $c_\infty$, on a circular boundary with a radius of 0.5, which encloses the growing precipitate. The initial solute concentration in the entire domain was set to $c_\infty$. For simplicity $P_e^{'} = 1$ or $\tau = \varepsilon^2/D$ and $\alpha = 0.5$ are used for all simulations. The grid resolution $N$ can be related to the model parameter $\varepsilon'$ through $N = L/\Delta x = 2/\varepsilon'$ where the characteristic length $L$ is chosen to be the size of the computational domain. With small parameter $\varepsilon' \to 0$ (or equivalently increasing grid resolution $N$), solution of phase-field equations (12)-(13) should converge to the sharp-interface solutions. Now the only free parameters in the model are the grid resolution $N$ (or $\varepsilon'$), the Damköhler number, $D_a$, and the far-field boundary condition, $c_\infty$. The total solid mass (or equivalently area $A$ occupied by the solid), and the radius of gyration $R_g = \int_A r dA \Big/ A$ (where $r$ is the distance of area element $dA$ from the origin) were recorded at various simulation times to study the precipitation kinetics.

The Laplacians in both phase-field and concentration equations were computed with the commonly used 5-point finite-difference stencil,

$$\nabla^2 \phi_{i,j} = \frac{\phi_{i+1,j} + \phi_{i-1,j} + \phi_{i,j+1} + \phi_{i,j-1} - 4\phi_{i,j}}{\Delta x^2}. \tag{16}$$

The magnitude of the gradient of $\phi$ was computed using the central difference scheme,

$$\left|\nabla \phi\right|_{i,j} = \frac{1}{\Delta x}\sqrt{\frac{\left(\phi_{i+1,j} - \phi_{i-1,j}\right)^2}{4} + \frac{\left(\phi_{i,j+1} - \phi_{i,j-1}\right)^2}{4}}, \tag{17}$$

and the curvature term in equation (12) was computed from the discretized version of the equation $\kappa' = \nabla \cdot \left(\nabla \phi / \left|\nabla \phi\right|\right)$, where

$$\nabla \cdot \left( \frac{\nabla \phi}{|\nabla \phi|} \right) = \frac{1}{\Delta x} \left\{ \begin{array}{c} \dfrac{\phi_{i+1,j} - \phi_{i,j}}{\sqrt{\left(\phi_{i+1,j} - \phi_{i,j}\right)^2 + \left(\phi_{i+1,j+1} + \phi_{i,j+1} - \phi_{i+1,j-1} - \phi_{i,j-1}\right)^2/16}} \\ -\dfrac{\phi_{i,j} - \phi_{i-1,j}}{\sqrt{\left(\phi_{i,j} - \phi_{i-1,j}\right)^2 + \left(\phi_{i-1,j+1} + \phi_{i,j+1} - \phi_{i-1,j-1} - \phi_{i,j-1}\right)^2/16}} \\ +\dfrac{\phi_{i,j+1} - \phi_{i,j}}{\sqrt{\left(\phi_{i,j+1} - \phi_{i,j}\right)^2 + \left(\phi_{i+1,j+1} + \phi_{i+1,j} - \phi_{i-1,j+1} - \phi_{i-1,j}\right)^2/16}} \\ -\dfrac{\phi_{i,j} - \phi_{i,j-1}}{\sqrt{\left(\phi_{i,j} - \phi_{i,j-1}\right)^2 + \left(\phi_{i+1,j-1} + \phi_{i+1,j} - \phi_{i-1,j-1} - \phi_{i-1,j}\right)^2/16}} \end{array} \right\}. \qquad (18)$$

The time discretization was implemented using the first order forward Euler method, and the time step satisfied the constraints from the stability conditions required for both the phase-field, $\phi$, and concentration field, $c$. The numerical implementation used for the two-dimensional simulations can be easily extended to three dimensional problems.

### III. ANALYTICAL SOLUTION FOR DIFFUSION-LIMITED PRECIPITATION

It is useful to first consider the precipitation problem with infinite Damköhler number (diffusion-limited) at a sharp circular solid-liquid interface since this can be formulated as a two-dimensional axisymmetric problem, solved analytically, and used to evaluate the phase-field model. The solid-liquid interface starts to advance into the region occupied by the fluid due to solute precipitation, and the corresponding dimensionless governing equations in polar coordinates obtained from Eqs. (6)-(8) are:

$$\frac{\partial c}{\partial t} = \frac{1}{P_e}\left(\frac{\partial^2 c}{\partial r^2} + \frac{1}{r}\frac{\partial c}{\partial r}\right), \qquad (19)$$

$$c\big|^+ = 0 \text{ on } \Gamma, \qquad (20)$$

$$v_s = \nabla c\big|^+ \cdot \mathbf{n}/2\alpha P_e \text{ on } \Gamma. \qquad (21)$$

The effect of the precipitation kinetics on the interface concentration disappears in Eq. (20) due to the infinite $D_a$. Suppose at time $t = 0$ the solid phase exists only at the origin, $r = 0$, and the far-field solute concentration boundary condition and initial condition are $c|_{r=+\infty, t>0} = c_\infty > 0$ and $c|_{r \neq 0, t=0} = c_\infty$, where $r$ is the radial coordinate, the analytical solutions to Eqs. (19)-(21) are:

$$c(r,t) = c_\infty - 2\alpha u_0 e^{u_0} E_i(u), \tag{22}$$

$$R(t) = 2\sqrt{u_0 t / P_e}, \tag{23}$$

$$A(t) = \pi R^2 = 4\pi u_0 t / P_e, \tag{24}$$

$$v_s = \partial R / \partial t = \sqrt{u_0 / (P_e t)}, \tag{25}$$

where $R$ is the radius of the solid-liquid interface and $A$ is the solid area at time $t$. The dimensionless variable $u$ is defined as $u = P_e r^2 / 4t$ ($r$ and $t$ are dimensionless radius and time). The exponential integral function $E_i$ is defined as $E_i(u) = \int_u^\infty e^{-t'}/t' \, dt'$. $u_0$ is the solution to the characteristic equation that is related to the far-field boundary concentration,

$$c_\infty = 2\alpha u_0 e^{u_0} E_i(u_0). \tag{26}$$

The right hand side of Eq. (26) is the Ivantsov function[26] in the theory of solidification. At the solid-liquid interface $u(r = R, t) = u_0$ is always satisfied, and $0 < c_\infty < 2\alpha$ must be satisfied for a solution of Eq. (26) to exist. Steady state growth (a constant growth velocity $v_s$) is not admissible for a circular interface with any far-field concentration $c_\infty$, in contrast to the growth of a one-dimensional planar interface where the growth velocity approaches a steady value with time.[17] The analytical solution reveals the classical scaling law, i.e. $A \propto t$, $R \propto t^{1/2}$ and $v_s \propto t^{-1/2}$. The solution is unstable with respect to any arbitrarily small perturbations on the shape of solid-liquid interface due to the Mullins-Sekerka instability.[2] This is especially true for low far-field concentration $c_\infty$, where an initially circular interface shape will eventually evolve into a complex dendritic structure.

The classical Witten-Sander diffusion-limited aggregation (DLA) model indicates that diffusion limited growth leads to the formation of a complex branched fractal pattern. It has been shown that a DLA cluster structure grows in time according to the scaling law,

$$R_g(t) \propto t^{1/(d_f - d + 2)}, \tag{27}$$

where $R_g$ is the mean radius (radius of gyration) of the cluster at growth time $t$. $d_f$ is the fractal dimensionality of the cluster and $d$ is the Euclidean dimensionality. Deutch and Meakin[27] derived this scaling law (Eq. (27)) subject to the conditions $d \geq 3$ by determining the steady state flux of material entering the cluster. First, the steady state concentration profile $c(r)$ can be found from the stationary diffusion equation in any dimension of $d$,

$$\frac{1}{r^{d-1}} \frac{d}{dr}\left(r^{d-1} \frac{d}{dr} c(r)\right) = 0, \tag{28}$$

where $r$ is the radial coordinate. By applying boundary conditions at a conceptual surface surrounding the cluster $c(r = R_g) = 0$ and far field concentration $c(r = \infty) = c_\infty$, the steady state concentration profile $c$ in the entire domain and the total solute diffusion flux $J$ at the conceptual surface can be easily shown to be

$$c(r) = c_\infty \left[1 - (R_g/r)^{d-2}\right] \tag{29}$$

$$J = Const \cdot R_g^{d-1} \left.\frac{dc}{dr}\right|_{r=R_g} = Const \cdot R_g^{d-2}. \tag{30}$$

for a dimensionality $d \geq 3$. Using the fractal dimensionality defined by the relationship $A \propto R_g^{d_f}$, and the relationship $J = dA/dt$ ($A$ is the total mass of the fractal), the scaling law given in Eq. (27) can be recovered from Eq. (30).

For solute precipitation problem, the diffusion-limited precipitation (as an analog of diffusion-limited aggregation) is achieved with an infinite Damköhler number, where the interface kinetics is neglected. Though there is no steady state concentration profile exists for $d = 2$, the total diffusion flux $J$ for $d = 2$ can be computed from the analytical concentration solution (Eq. (22)) to be $J = 2\pi R \left.\frac{dc}{dr}\right|_{r=R} = 8\pi\alpha u_0 = Const$. Therefore Eq. (30) is still valid for $d = 2$ and the scaling law (Eq.(27)) is then extended to all embedding dimensionality greater than or equal 2, namely the 2D simulations of the present study.

## IV. RESULTS AND DISCUSSION

Instead of the original Witten and Sander's algorithm[4] of growing a DLA cluster by adding particles launched from infinity, one at a time, to an initial seed or growth site, Meakin and Deutch designed a new algorithm where an initial seed is immersed in a low density cloud of randomly walking particles that stick either to the seed or part of the cluster.[20] The cluster that grows from the seed is formed as a random dendritic cluster at first, very similar to clusters generated by the original DLA model. The growing cluster consumes the randomly walking particles in its vicinity and cluster density decreases until the density of the cluster approaches the density of the cloud of particles in which it is growing.[20] It was observed that for a very low initial particle concentration (the classical Witten-Sander model corresponds to the limit of vanishing diffusing particle concentration), the resulting cluster has a fractal dimensionality of $d_f = 5d/6$. Specifically, the radius of gyration and the total mass increase according to the scaling laws $R_g(t) \propto t^{1/(d_f - d + 2)}$ (Eq. (27)) and $A(t) \propto t$. The effective fractal dimensionality is the same as that of clusters generated by the original DLA model that depends on the underlying lattice. In two-dimensional space, the fractal dimension decreases with decreasing number of growth directions from $d_f \approx 1.71$[28] for off-lattice DLA to $d_f \approx 1.6$ for growth in three directions on a square lattice, the effective fractal dimension is $\approx 1.64$ for clusters of $4 \times 10^6$ sites.[29]

For a finite initial particle concentration, the average density of the growing cluster is much larger than the average density of the solute at the early stages of growth and the growth of the dendrite is indistinguishable to the growth of a classical DLA cluster (addition of one particle at a time, without other particles present corresponding to $c \to 0$ limit). As the dendritic pattern increases in size, its density decreases and when the density of the dendritic pattern approaches the solute density, a crossover in the structure and growth kinetics[20, 30, 31] can be observed from fractal growth to compact aggregate growth that follows scaling laws $R_g(t) \propto t$ and $A(t) \propto t^d$. The critical cluster size $R_g = \xi$ of crossover is given by $\xi \sim c_\infty^{-1/(d-d_f)}$.[31]

The phase-field equations (12)-(14) can be used to solve the corresponding sharp-interface equations (6)-(8), which is a continuum analogue of a DLA cluster grows from a medium with an initial finite concentration $c_\infty$ of particles. The phase-field simulations were performed on a finite difference grid with a resolution of $N = 1000$ or equivalently, $\varepsilon' = 0.002$. The simulations starts with a uniform concentrations, $c_\infty$, everywhere. A circular boundary where the solute concentration is fixed at $c_\infty$ is used to mimic the far-field fixed solute concentration. A large Damköhler number, $D_a$, can be used to represent the fast reaction compared to the solute diffusion. Different values of $c_\infty = 0.1$, 0.5 and 1.0 were used to investigate the effect of solution concentration on the cluster growth. Figures 2, 3 and 4 show snapshots of growth patterns at various simulation times $t$ with different $c_\infty$. At low concentration (Fig.2), the interfacial patterns generated cannot be exactly the same with the classical particle Monte Carlo simulations. Those patterns posses symmetry as a result of grid-based continuum simulations, while the general morphology still resembles the classical Witten-Sander cluster. The branched growth is evidenced as branches dies due to the competition and the screen effect between branches and their neighbors and new branches are created via tip-splitting processes. A compact aggregates behavior was observed at larger solute concentrations, as shown in Figs. 3 and 4.

The phase-field simulations generate branched patterns that are in some ways resemble DLA clusters, and they might become more similar to much larger patterns (patterns with more generations of branching). However, there are important differences between the patterns and the algorithms that generated them due to the way in which disorder enters into the two models. In the DLA model the random addition of particles corresponds to growth fluctuations that vary in both space and time, and because of the spatial fluctuations the symmetry of the underlying lattice is destroyed (it may be recovered on very large scales due to averaging of the spatial fluctuations). In the phase-field model, the growth fluctuations come about as a result of "quenched disorder" that comes about as a result of structure in the "substrate" in which the solid is growing. In both cases, the effects of the growth fluctuations are amplified due to the Mullins Sekerka instability, and this amplification is responsible for the rapid development of a highly branched structure.

Because the growth perturbation in the phase-field model retains the four-fold symmetry of the lattice, the resulting branched pattern also retains the four-fold symmetry. Figure 5 shows an experimental analog,[32] where the interfacial motion of two incompressible fluids with different viscosities in a Hele-Shaw cell was studies. It is well known that a similarity exists between above interfacial motion and the diffusion-limited aggregation, both of which are described by the same set of equations.[21]  In, Fig 5, dyed oil was injected through a small hole in one wall of a Hele Shaw cell (two parallel glass plates with a small gap between them) filled with glycerol. A shallow rectangular network of grooves had been etched in on interior wall of the cell, and this creates a similar symmetry as our phase-field simulation that influences the propagation of the interface between the two immiscible fluids. The viscosity of the glycerol is much larger than the viscosity of the oil, and for sufficiently large injection rates, viscous forces dominate capillary forces, and the fluid-fluid displacement processes in a smooth walled Hele-Shaw cell can be described by a diffusion-limited aggregation equation (the pressure field in the viscous fluid plays the same role as the solute concentration field).

The phase-field equations describe diffusion-limited precipitation in a homogeneous field, and the analytical solution for an initial state that has two concentric circular boundaries predicts shape preserving growth of the inner boundary. The Mullins Sekerka instability makes the growth process exquisitely sensitive to perturbation (they grow exponentially with increasing time) and in the simulation the discretization of the growth equations on a square lattice provides the perturbation. In this respect, branched patters are generated in the simulations because the circular boundary cannot be perfectly represented by a discretization on a square lattice. The experimental pattern shown in Fig. 5 does not perfectly preserve the nominal four-fold symmetry. This is not surprising since small fabrication errors and contamination by dust, grease etc. is very difficult to avoid.

Figures 6 and 7 show the growth of the total area $A$ and radius of gyration $R_g$ with increasing simulation time $t$ in the form of log-log plot for an initial and far-field concentrations $c_\infty$ = 0.1, 0.5 and 1. As expected, the scaling laws $A(t) \propto t$ and $R_g(t) \propto t^{1/(d_f - d + 2)}$ are observed for low solute concentration $c_\infty$ = 0.1, where $d_f$ is measured to be 1.68. The density of the cluster structure decreases with increasing cluster size and a crossover can be observed when the cluster density approaches the bath solute concentration $c_\infty$. With increasing solute concentration $c_\infty$, the crossover critical length scale, $\xi$, decreases and scaling laws $A \propto t^2$ and $R_g \propto t$ are observed at larger solute concentration, $c_\infty$, where the cluster structure generated becomes more compact with little open space. To determine the fractal dimensionality, $d_f$, of the growth patterns in the limit of low solute concentration, the dependence of the logarithm of the area, $A$, covered by the solid, on both the radius of gyration, $R_g$, and the maximum cluster radius, $R_{max}$, are presented in Fig. 8 for a low concentration $c_\infty = 0.1$. Two distinct regions can be identified with slopes of 2.0 and $d_f$, corresponding to the period before the perturbations are introduced and become substantial enough to broke circular symmetry and the period after the circular symmetry has been broken due to the Mullins-Sekerka instability.[2] The perturbation in the phase-field simulations was initially very small and grows fast with simulation time, in contrast to the DLA models, where the large perturbation is introduced right at the initial stage of the simulation. The fractal dimensionality is estimated to be $d_f = 1.68$ in the second period of fractal growth. This number is close to the fractal dimensionality of $d_f \approx 1.64$ that obtained for large square lattice DLA clusters.

In addition to the growth from a solid seed (quarter disk), it is also very interesting to examine the inverse situation, where the liquid is surrounded by solid and growing dendrites from the internal surface of solid. An initial concentration of $c_o$ is used in the liquid and the concentration at the inner boundary is $c|_{r=0, t\geq 0} = c_0 > 0$. This illustrates dendritic growth from multiple sites and the competition for solute among various braches leading to screening effects and the growth of large branches at the expense of their smaller neighbors, which eventually stop growing altogether. Figures 9 and 10 show snapshots of these patterns at various simulation time $t$ for two concentrations $c_o = 0.5$ and 0.75, respectively. As expected, a more compact solid structure is generated with a higher concentration.

## V. CONCLUSIONS

A phase-field approach to model solute precipitation and dissolution at solid-liquid interfaces has been applied to study diffusion-limited precipitation by eliminating the details of interfacial kinetics (the chemical reaction at the interface is not relevant in this study providing that it is large enough to ensure diffusion-limited precipitation conditions at all times or stages of growth). The validation tests show a good agreement with the available analytical solution. The subsequent study of dendritic growth due to diffusion-limited precipitation yields a solid-liquid interfacial pattern with fractal geometry. The fractal dimensionality for low solute concentration was estimated to be 1.68, very close to the fractal dimensionality of 1.64 estimated for large square lattice DLA clusters.

**ACKNOWLEDGMENTS**

This work was supported by the U.S. Department of Energy, Office of Science Scientific Discovery through Advanced Computing Program. The Idaho National Laboratory is operated for the U.S. Department of Energy by the Battelle Energy Alliance under Contract DE-AC07-05ID14517.

FIG. 1. Growth velocity of perturbations as a function of wavenumber. The grey line indicates the prediction of the Mullins-Sekerka linear stability analysis. The dashed line shows the behavior with a zero capillary length and a lattice cutoff.

FIG. 2. Snapshots of interfacial patterns at various simulation time $t$ for far-field concentration $c_\infty = 0.1$.

FIG. 3. Snapshots of interfacial patterns at various simulation time $t$ for far-field concentration $c_\infty = 0.5$.

FIG. 4. Snapshots of interfacial patterns at various simulation time $t$ for far-field concentration $c_\infty = 1.0$.

FIG. 5. Experimental radial viscous fingering pattern in Hele-Shaw cells.

FIG. 6. Logarithm plot of the total solid area $A$ varying with simulation time $t$ for various far-field concentration $c_\infty = 0.1$, 0.5 and 1.0.

FIG. 7. Logarithm plot of radius of gyration $R_g$ varying with simulation time $t$ for various far-field concentration $c_\infty = 0.1$, 0.5 and 1.0.

FIG. 8. Logarithm plot of the total area $A$ varying with radius of gyration $R_g$ and maximum radius $R_{max}$ for far-field concentration $c_\infty = 0.1$.

FIG. 9. Snapshots of interfacial patterns at various simulation time $t$ for precipitation inside a solid pipe with a fixed center concentration $c_o = 0.5$.

FIG. 10. Snapshots of interfacial patterns at various simulation time $t$ for precipitation inside a solid pipe with a fixed center concentration $c_o = 0.75$.

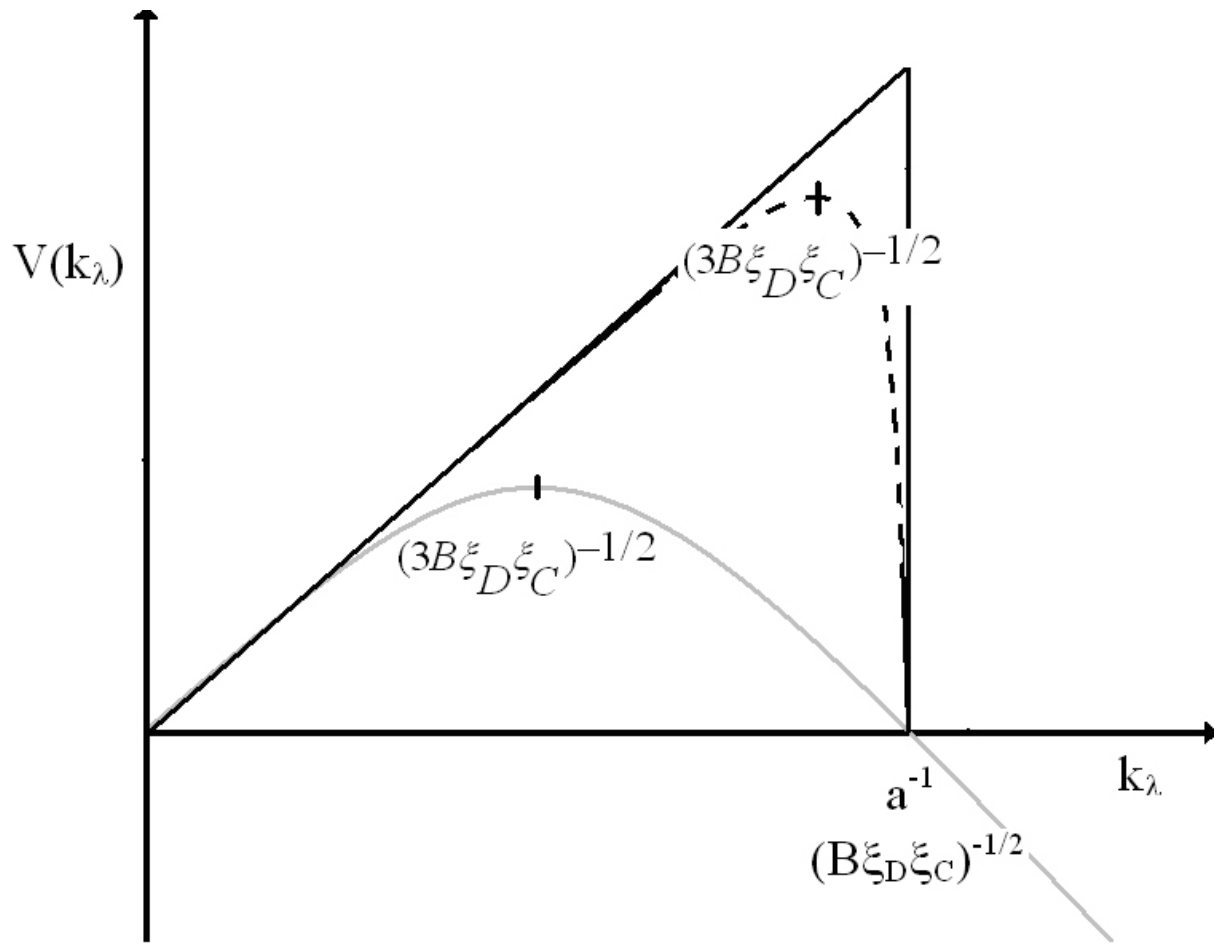

| a) t=0.04 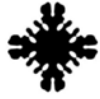 | b) t=0.08 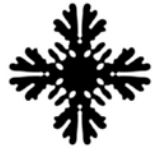 |
| c) t=0.12 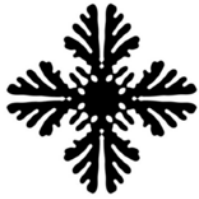 | d) t=0.16 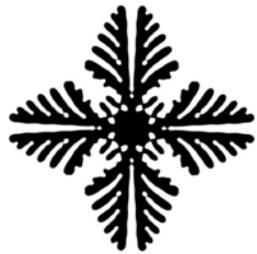 |

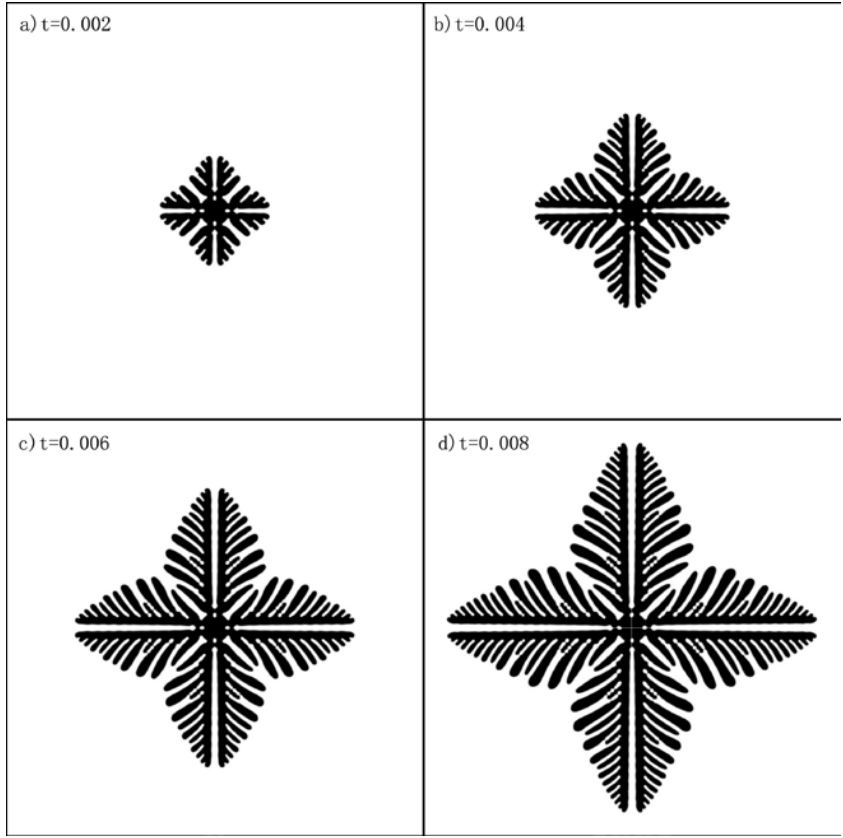

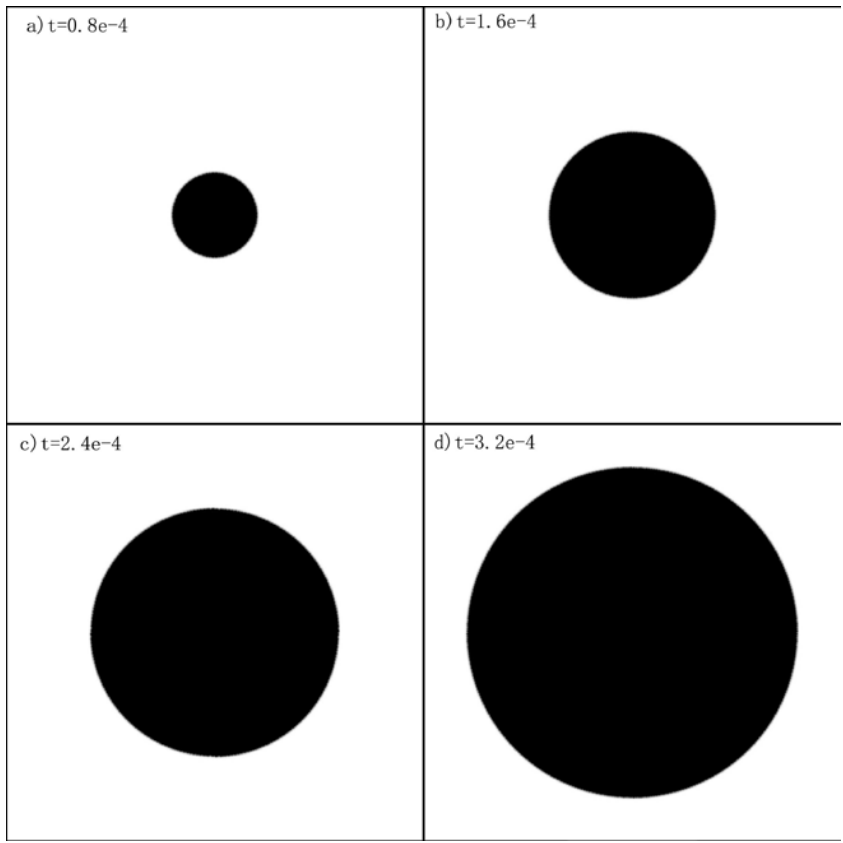

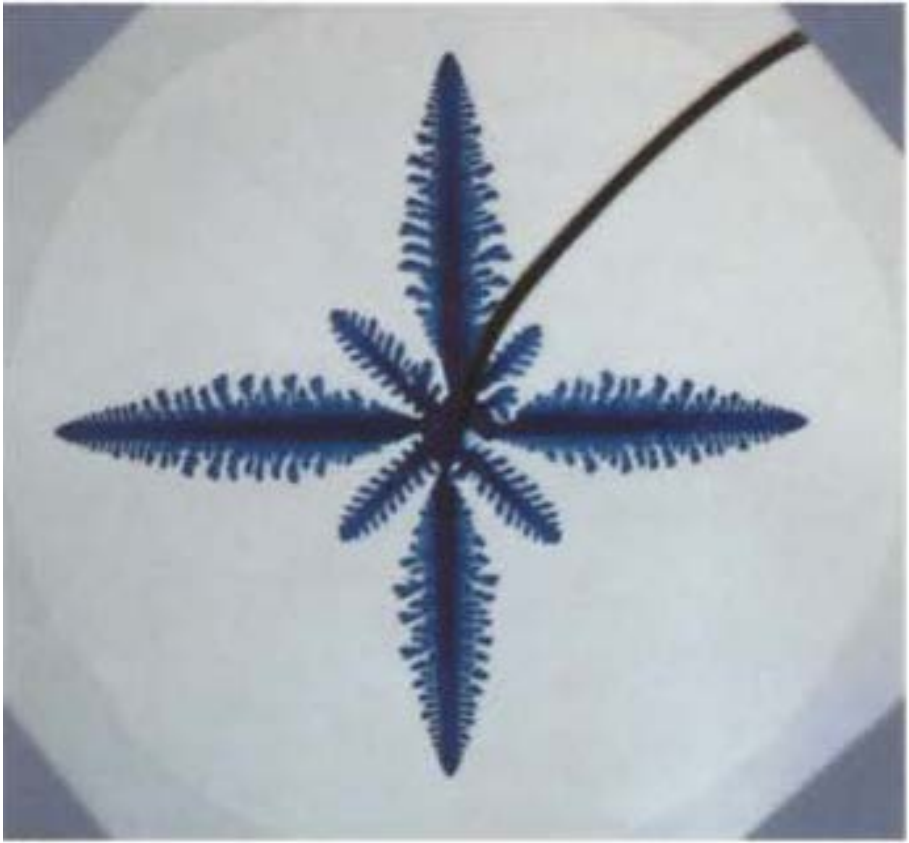

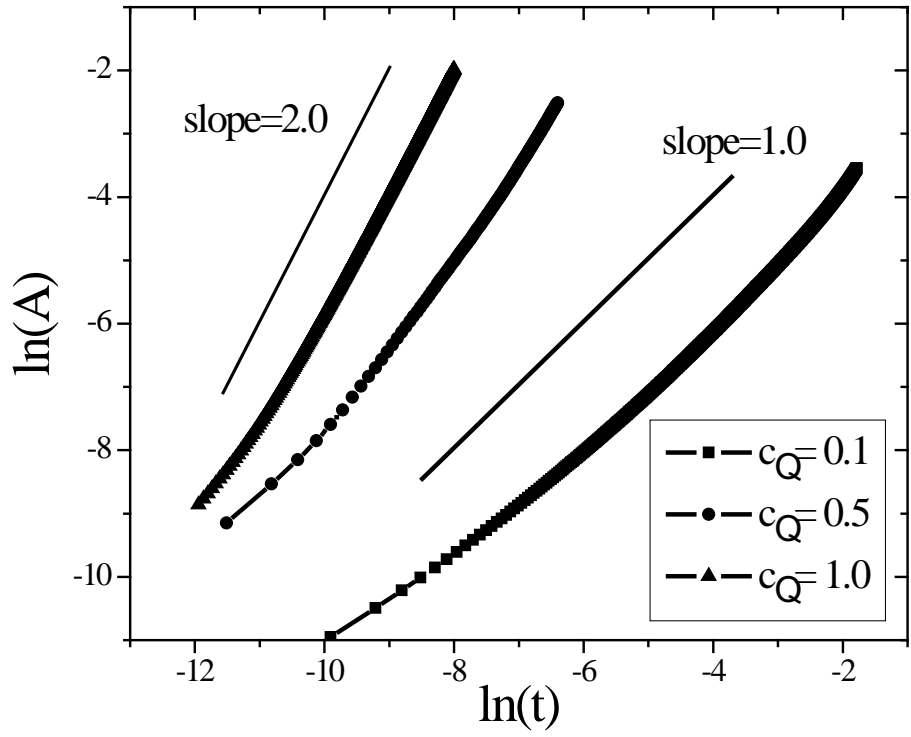

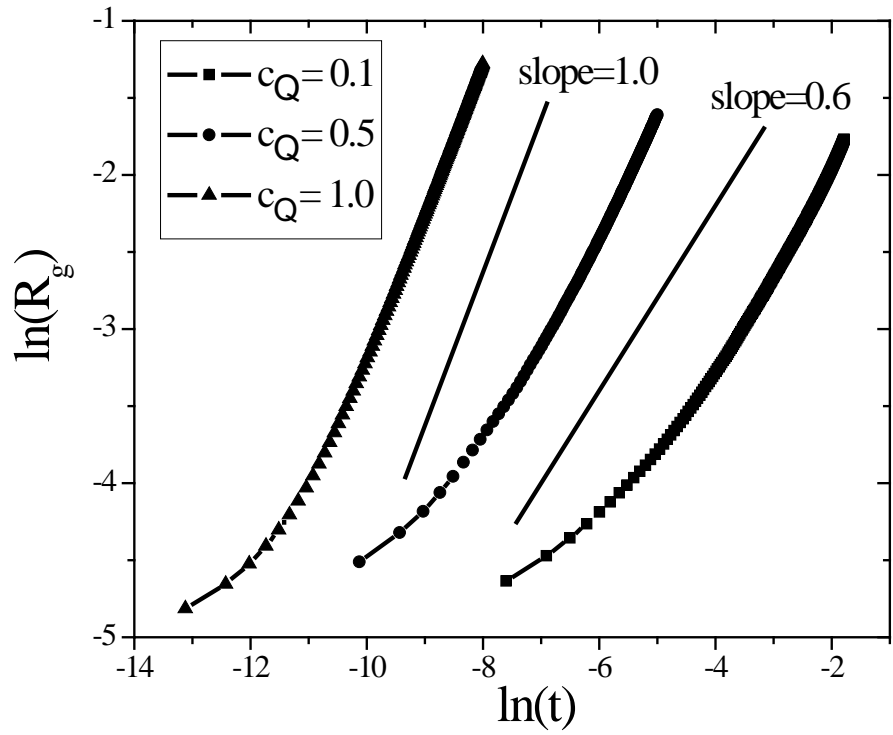

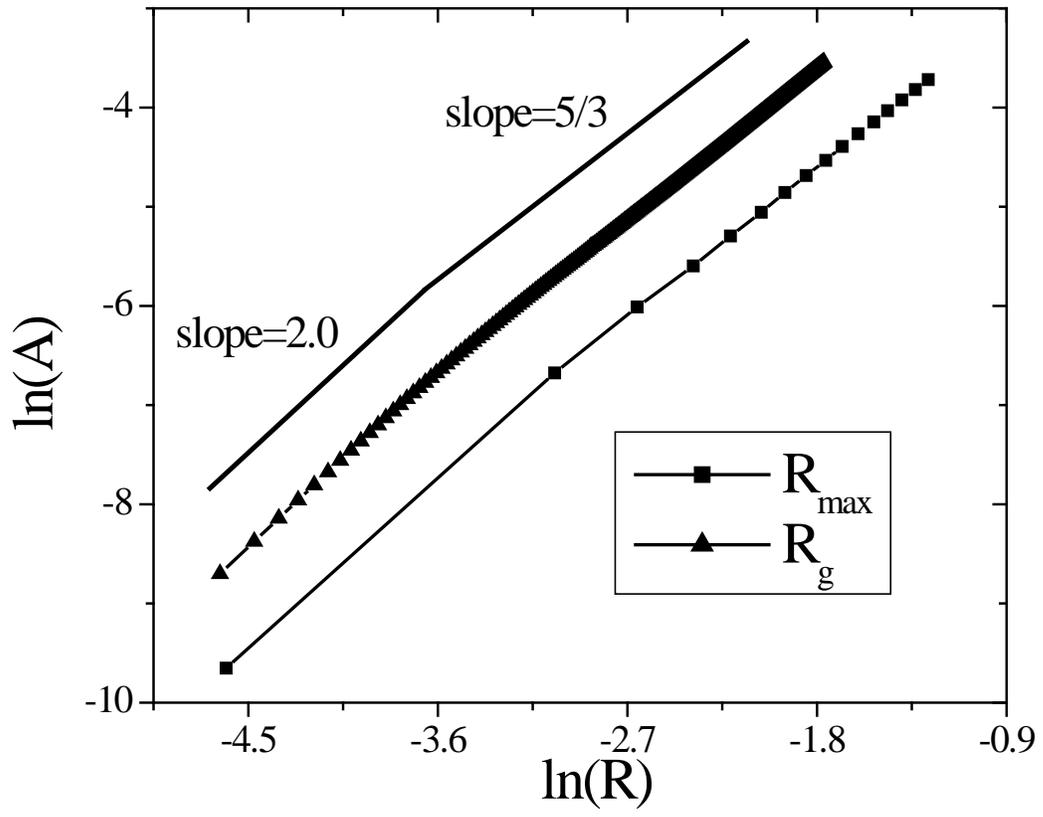

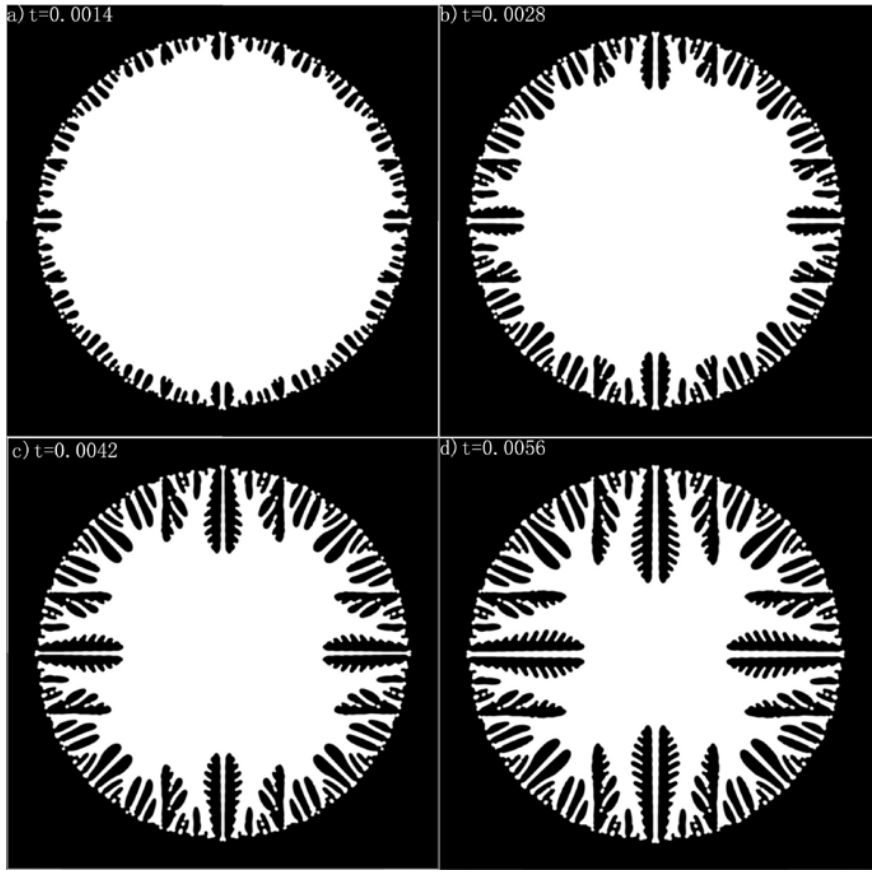

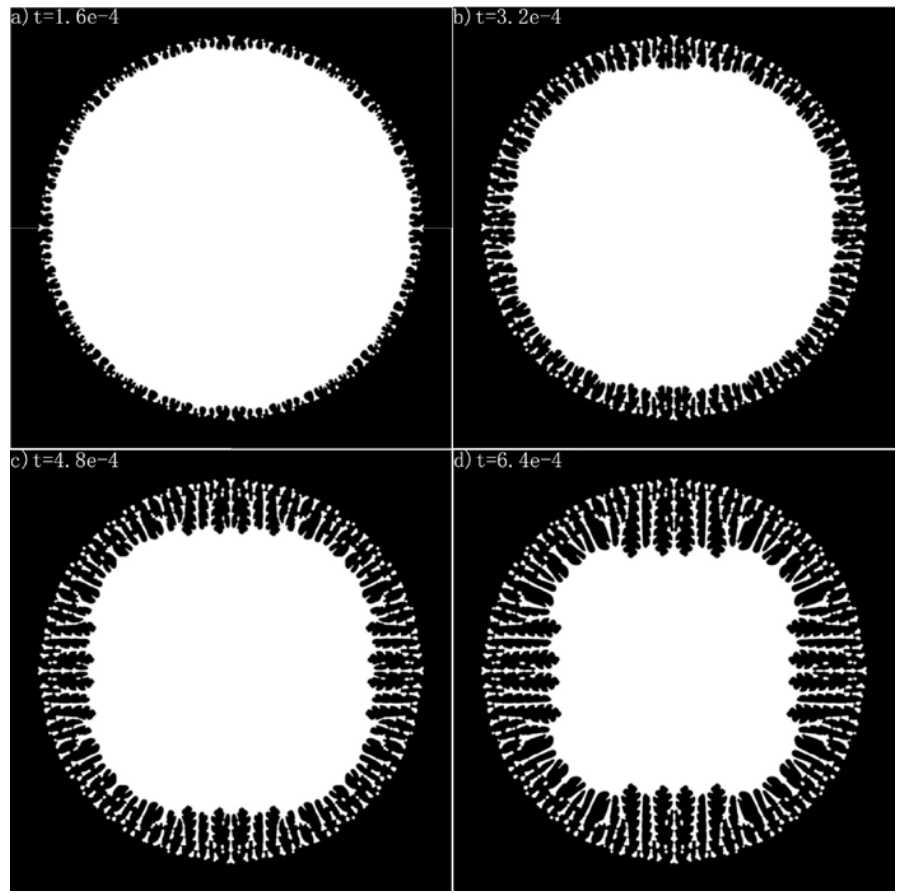33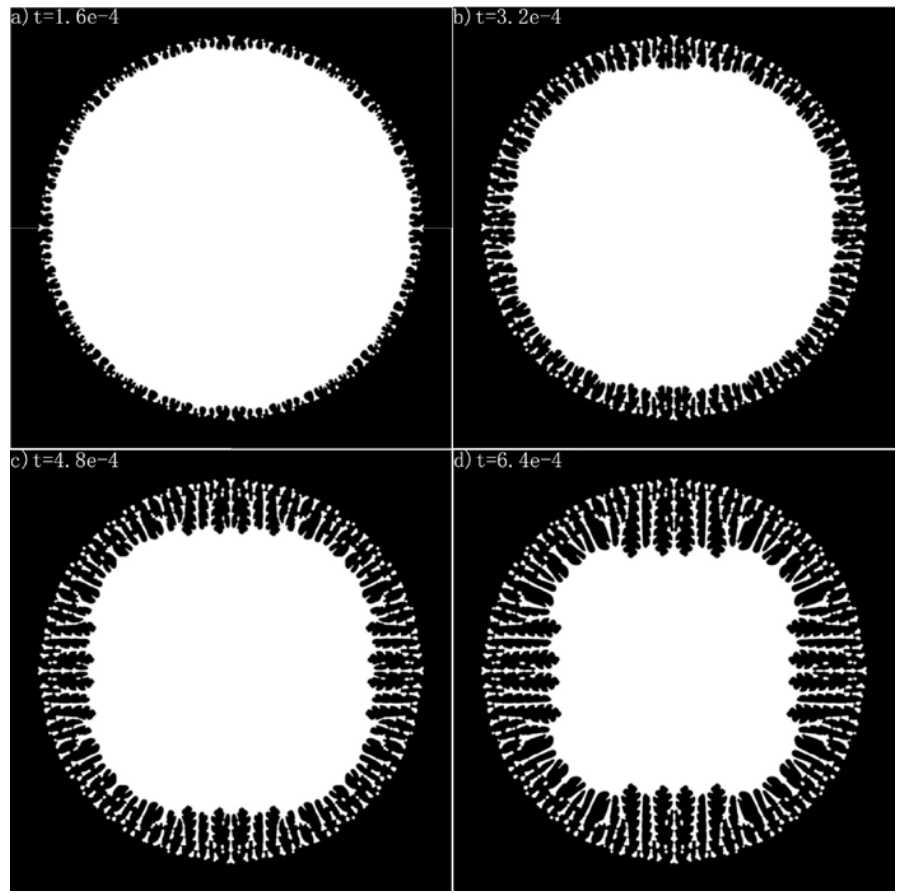